\documentclass[12pt]{iopart}
\usepackage{graphicx}
\usepackage{cite}
\begin{document}

\title{Identifying valence structure in LiFeAs and NaFeAs with core-level spectroscopy}

% authors list
\author{E Z Kurmaev$^1$, J A McLeod$^2$, N A Skorikov$^1$, L D Finkelstein$^1$, A Moewes$^2$, Yu A Izyumov$^1$, and S Clarke$^3$}
\address{$^1$ Institute of Metal Physics, Russian Academy of Sciences-Ural Division, 620219 Yekaterinburg, Russia}
\address{$^2$ Department of Physics and Engineering Physics, University of Saskatchewan, 116 Science Place, Saskatoon, Saskatchewan S7N 5E2, Canada}
\ead{john.mcleod@usask.ca}
\address{$^3$ Department of Chemistry, University of Oxford, Inorganic Chemistry Laboratory, South Parks Road, Oxford, OX1 3QR, United Kingdom}

\begin{abstract}
Resonant X-ray emission spectroscopy (XES) measurements at Fe \textit{L}$_{\mathit{2,3}}$-edges and electronic structure calculations of LiFeAs and NaFeAs are presented. Experiment and theory show that in the vicinity of the Fermi energy, the density of states is dominated by contributions from Fe \textit{3d}-states. The comparison of Fe \textit{L}$_{\mathit{2,3}}$ XES with spectra from related FeAs-compounds reveal similar trends in energy and the ratio of the intensity of the \textit{L}$_{\mathit{2}}$ and \textit{L}$_{\mathit{3}}$ peaks (I(\textit{L}$_{\mathit{2}}$)/I(\textit{L}$_{\mathit{3}}$) ratio). The I(\textit{L}$_{\mathit{2}}$)/I(\textit{L}$_{\mathit{3}}$) ratio of all FeAs-based superconductors is found to be closer to that of metallic Fe than that of the strongly-correlated FeO. We conclude that iron-based superconductors are weakly- or at most moderately-correlated systems.
\end{abstract}

% suggested PACS numbers
\pacs{71.20.Dg, 74.25.Jb}
\submitto{\JPCM}
\maketitle

\section{Introduction}
Recently, a new class of iron-based superconductors was discovered. LiFeAs and NaFeAs were found to be superconducting with T$_c$ = 18K and 9K, respectively \cite{pitcher08, tapp08, wang08, parker08}. These compounds have an unexpected lack in magnetic order at all temperatures whereas in \textit{RE}OFeAs (where \textit{RE} is a rare earth element) and \textit{AE}Fe$_2$As$_2$ (where \textit{AE} is an alkaline earth element) the Fe magnetic moments adopt a collinear antiferromagnetic (c-AFM) order at low temperatures \cite{tapp08}. In contrast with other iron-based compounds, no further doping is necessary to induce superconductivity and the spin-density wave (SDW) state appears to be notably absent from these new systems \cite{chu09}. This suggests LiFeAs and NaFeAs can offer important insights to understanding the mechanism of superconductivity in iron-based superconductors. The absence of SDW transitions and the relatively low T$_c$ in comparison with \textit{RE}OFeAs and \textit{AE}Fe$_2$As$_2$ iron-based superconductors make these two compounds possible candidates for conventional BCS superconductors \cite{jishi08}. However the estimation of the electron-phonon coupling parameter $\lambda$ from band structure calculations \cite{jishi08} gives low values of $\lambda$ = 0.29 and 0.27 for LiFeAs and NaFeAs respectively, which are too weak to account for conventional BCS superconductivity in this class of superconductors. Experimental studies of electronic structure of LiFeAs and NaFeAs superconductors and their comparison with that of \textit{RE}OFeAs and \textit{AE}Fe$_2$As$_2$ iron-based superconductors are important in resolving this puzzle. In this manuscript we present and analyze X-ray emission spectroscopy (XES) measurements at the Fe \textit{L}$_{\mathit{2,3}}$-edge of  LiFeAs and NaFeAs. These measurements probe the occupied Fe \textit{3d} density of states (DOS). We compare the measured spectra with our electronic structure calculations of the valence structure. We contrast our findings with our previous studies of \textit{RE}OFeAs (\textit{RE} = La, Sm) \cite{kurmaev08} and CaFe$_2$As$_2$ \cite{kurmaev09}.

\section{Experimental and Calculation Details}
LiFeAs and NaFeAs were synthesized by the reaction of stoichiometric  quantities of elemental Li (or Na), Fe and As. Fe and As powders were ground together and added to pieces of Li (or Na) in a tantalum tube, which was then sealed by welding under 1 atm argon gas; the mixture was heated to 800 $^\circ$C for 2 days. Preliminary characterization of the resulting product by powder X-ray diffraction revealed pure LiFeAs and NaFeAs phases. For details of sample preparation see \cite{pitcher08} and \cite{parker08}.

The X-ray emission measurements of LiFeAs and NaFeAs were performed at the soft X-ray fluorescence endstation at Beamline 8.0.1 of the Advanced Light Source at Lawrence Berkeley National Laboratory \cite{jia95}. The endstation uses a Rowland circle geometry X-ray spectrometer with spherical gratings and an area sensitive multichannel detector. We have measured resonant and non-resonant Fe \textit{L}$_{\mathit{2,3}}$ (\textit{3d4s} $\rightarrow$ \textit{2p} transition) X-ray emission spectra (XES). The instrumental resolving power (E/$\Delta$E) for Fe \textit{L}$_{\mathit{2,3}}$ spectra was 10$^{3}$. All spectra were normalized to the incident photon current using a highly transparent gold mesh in front of the sample to correct for intensity fluctuations in the photon beam. The excitations for the XES measurements were determined from Fe \textit{2p} X-ray absorption spectroscopy (XAS) measurements (using total electron yield mode); the chosen energies corresponded to the location of the \textit{L}$_{\mathit{2}}$ and \textit{L}$_{\mathit{3}}$ thresholds, an energy between them, and an energy well above the \textit{L}$_\mathit{2}$ threshold. 
 
All density of states calculations were performed within the full-potential linear augmented plane-wave method as implemented in WIEN2k code \cite{blaha01}. For the exchange-correlation potential we used the generalized gradient approximation in the Perdew-Burke-Ernzerhof variant \cite{perdew96}. The Brillouin zone integrations were performed with a 12 $\times$ 12 $\times$ 7 special \textit{k}-point grid for LiFeAs and a 12 $\times$ 12 $\times$ 6 special \textit{k}-point grid for NaFeAs. and \textit{R}$_{MT}^{min}$\textit{K}$_{max}$ = 7 (the product of the smallest of the atomic sphere radii \textit{R}$_{MT}$ and the plane wave cutoff parameter \textit{K}$_{max}$) was used for the expansion of the basis set. The experimental values of the high-temperature lattice constants and atomic positions are used \cite{pitcher08, parker08}. Both compounds have a tetragonal crystal lattice in the P4/nmm space group. In NaFeAs, the lattice constants are \textit{a} = 3.9494 $\mathrm{\AA}$ and \textit{c} = 7.0396 $\mathrm{\AA}$, whereas in LiFeAs, \textit{a} = 3.7754 $\mathrm{\AA}$ and \textit{c} = 6.3534 $\mathrm{\AA}$ (for a summary of structural parameters see Table \ref{tbl:lattice}). The atomic sphere radii were chosen as \textit{R}$_{Na}$ = 2.5, \textit{R}$_{Fe}$ = 2.3, \textit{R}$_{As}$ =2 .04 a.u. and \textit{R}$_{Li}$ = 2.36, \textit{R}$_{Fe}$ = 2.28, \textit{R}$_{As}$ = 2.02 a.u. for NaFeAs and LiFeAs respectively. The sphere radii were chosen in such a way that the spheres were nearly touching.
\begin{table}
\begin{tabular}{cccccc}
Compound & Fe-Fe & NN(Fe) & Fe-R & NN(R) & Pt. Grp.\\
\hline
LaOFeAs \cite{chen08} & 2.85 & 4 & 2.41 & 4 & \textit{-42m}\\
CaFe$_2$As$_2$ \cite{ronning08} & 2.74 & 4 & 2.32 & 4 & \textit{-42m}\\
NaFeAs & 2.79 & 4 & 2.44 & 4 & \textit{-42m}\\
LiFeAs & 2.67 & 4 & 2.41 & 4 & \textit{-42m}\\
Fe (bcc) \cite{basinski55} & 2.54 & 8 & - & - & \textit{m3m}\\
FeO \cite{fjellvag96} & 3.06 & 12 & 2.16 & 6 & \textit{m3m}\\
\end{tabular}
\caption{\label{tbl:lattice}Structural parameters for various Fe compounds. Here ``Fe-Fe'' refers to the distance between iron atoms, ``Fe-R'' refers to the distance between the iron atom and the anion, both in $\mathrm{\AA}$. NN() refers to the number of nearest neighbours of the specified type. The final column, ``Pt. Grp.'' refers to the point group around the Fe atom.}
\end{table}

\section{Results and Discussion}
\begin{figure}
\includegraphics[width=3in]{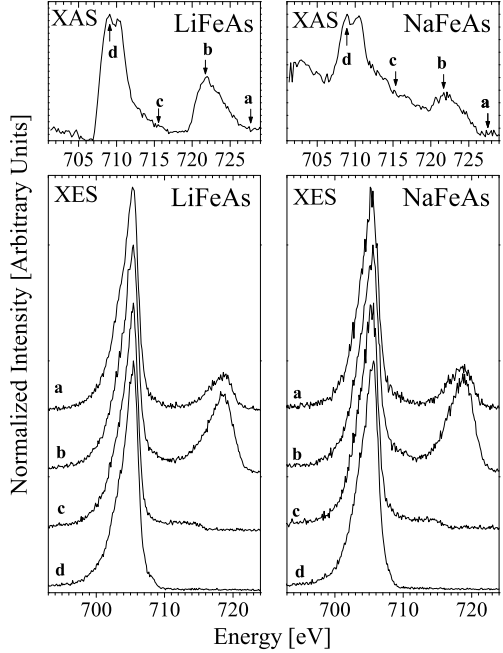}
\caption{\label{fig:xes_summary}Summary of spectra for LiFeAs (left side) and NaFeAs (right side). The excitation energies for resonant Fe \textit{L}$_{\mathit{2,3}}$ XES are indicated by arrows in the XAS spectra in the left panels.}
\end{figure}

The measured XES and XAS spectra are shown in Figure \ref{fig:xes_summary}. The LiFeAs and NaFeAs Fe \textit{L}$_{\mathit{2,3}}$ XES (Fig. \ref{fig:xes_summary}, bottom panels) indicate two main bands located around 705 and 718 eV, these correspond to the Fe \textit{L}$_{\mathit{3}}$ ($3d4s \rightarrow 2p_{3/2}$ transitions) and Fe \textit{L}$_{\mathit{2}}$ ($3d4s \rightarrow 2p_{1/2}$ transitions) normal emission lines separated by the spin-orbital splitting of Fe \textit{2p}-states. The non-resonant Fe \textit{L}$_{\mathit{2,3}}$ XES (XES curve \textbf{a}) for both materials lacks the low-energy satellite structure typical for correlated systems (for instance for FeO \cite{galakhov97}) and the main peak is sharp and similar to metallic iron \cite{gao06}. The resonant XES spectra (XES curves \textbf{b, c, d} in the bottom panels of Figure \ref{fig:xes_summary}) show no energy-loss features; this indicates that even resonant Fe \textit{L}$_{\mathit{2,3}}$ XES probes mainly the partial DOS in these materials. The XAS spectra (top panels of Figure \ref{fig:xes_summary}) indicate the absorption thresholds for Fe \textit{2p}$_{\mathit{1/2}}$ and \textit{2p}$_{\mathit{3/2}}$ electrons, and are used to determine the appropriate excitation energies for resonant XES. The detailed features in XAS spectra reveal information about the Fe \textit{3d} unoccupied states (the conduction band), however due to the greater effective atomic potential when a \textit{2p} core electron is removed the states in a XAS spectra are greatly distorted from those in the unperturbed crystal; this is referred to as the ``core-hole effect''.

\begin{figure}
\includegraphics[width=3in]{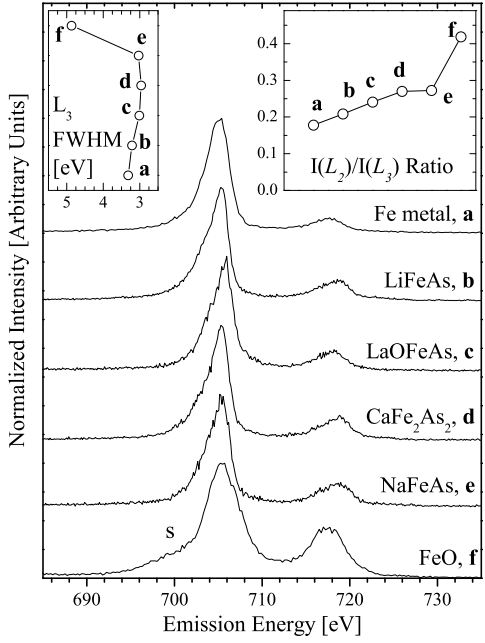}
\caption{\label{fig:l2l3}Comparison of the \textit{L}$_{\mathit{2,3}}$ XES spectra for metallic Fe, LiFeAs, LaOFeAs\cite{kurmaev08}, CaFe$_2$As$_2$\cite{kurmaev09}, NaFeAs, and FeO. Note the low energy shoulder \textbf{s} appears in FeO but not in any of the other spectra. The left inset shows the full-width at half maximum (FWHM) of the Fe \textit{L}$_{\mathit{3}}$ peak. The right inset shows the I(\textit{L}$_{\mathit{2}}$)/I(\textit{L}$_{\mathit{3}}$) ratios for metallic Fe, LaOFeAs, CaFe$_2$As$_2$, and correlated FeO. The I(\textit{L}$_{\mathit{2}}$)/I(\textit{L}$_{\mathit{3}}$) ratio was calculated from the ratios of the integral under the \textit{L}$_{\mathit{2}}$ and \textit{L}$_{\mathit{3}}$ peaks, respectively.}
\end{figure}

The integral of the \textit{L}$_{\mathit{2}}$ and \textit{L}$_{\mathit{3}}$ peaks in a non-resonant XES measurement are related to the population of the \textit{2p}$_\mathrm{1/2}$ and \textit{2p}$_\mathrm{3/2}$ states respectively. For free atoms the ratio of the integrals of the \textit{L}$_{\mathit{2}}$ and \textit{L}$_{\mathit{3}}$ XES peaks (the I(\textit{L}$_{\mathit{2}}$)/I(\textit{L}$_{\mathit{3}}$) ratio) should be equal to $\frac{1}{2}$. In metals the radiationless \textit{L}$_2$\textit{L}$_3$\textit{M}$_{4,5}$ Coster-Kronig (C-K) transitions strongly reduce the I(\textit{L}$_{\mathit{2}}$)/I(\textit{L}$_{\mathit{3}}$) ratio \cite{kurmaev05}. Figure \ref{fig:l2l3} illustrates that the I(\textit{L}$_{\mathit{2}}$)/I(\textit{L}$_{\mathit{3}}$) ratio is almost identical for all previously mentioned FeAs-compounds. This ratio is closer to that of metallic Fe than to that of FeO, and indicates that the Fe \textit{3d} states in FeAs-compounds are much less correlated then those in FeO. Further, the low energy edge of the \textit{L}$_{\mathit{3}}$ peaks in the FeAs-compounds and metallic Fe lack the prominent satellite that is present in FeO. As previously mentioned this also supports the conclusion that the Fe \textit{3d} electrons in FeAs-compounds are largely itinerant. For metallic compounds, the Fe \textit{3d} bandwidth should decrease with increasing Fe --- Fe distance and decreasing number of Fe --- Fe neighbours (see Table \ref{tbl:lattice}), and this is somewhat demonstrated in the full-width at half maximum (FWHM) of the Fe \textit{L}$_{\mathit{3}}$ peak (Figure \ref{fig:l2l3}, top left inset), which reduces from $\sim$3.3 eV for Fe metal to $\sim$3.0 eV for NaFeAs. Except for LaOFeAs, which we would expect to have a smaller FWHM than NaFeAs, the trend in decreasing FWHM qualitatively matches the trend in increasing Fe --- Fe distance in Table \ref{tbl:lattice} for the metallic compounds. It should be noted that these changes in FWHM are less then the instrumental resolution by roughly a factor of two and that core-level spectroscopy is not an appropriate technique to probe this effect -- so we do not attempt to draw any conclusions from this trend. There is, however, a clear difference in the FWHM between the FeAs-compounds and FeO. 

\begin{figure}
\includegraphics[width=3in]{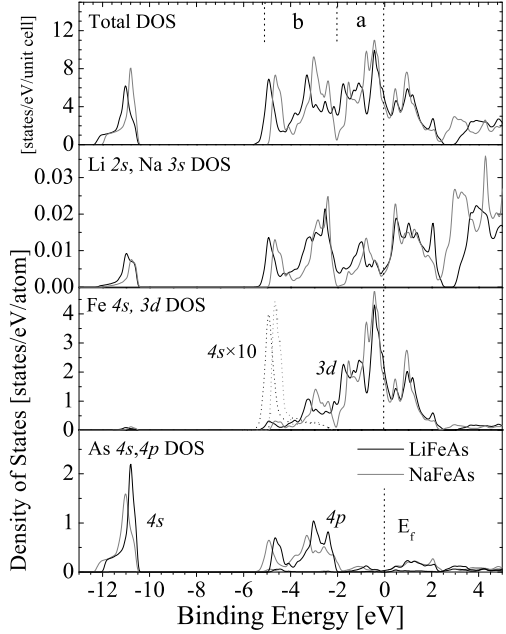}
\caption{\label{fig:dos}Calculated DOS for LiFeAs and NaFeAs. The dotted lines in the Fe \textit{3s,3p} DOS plot refer to the \textit{3s} states increased by a factor of 10. The As \textit{4s} states are separated from the \textit{4p} states so they are plotted with the same line style. The y-axis in the total DOS plot is in units of [states/eV/unit cell]. Two regions, \textbf{a} and \textbf{b} are identified in the total DOS plot. Fe \textit{3d}-states dominate in region \textbf{a}. Roughly even contribution from Fe \textbf{3d}-states and As \textit{4p}-states occurs in region \textbf{b}.}
\end{figure}

The calculated electronic structure of LiFeAs and NaFeAs are presented in Figure \ref{fig:dos}. Our calculations match those performed in Reference  \cite{jishi08}. The calculated partial density of states (DOS) distribution is quite similar for both compounds. In particular, the density of states in the vicinity of the Fermi level (0 to -2 eV)  is dominated by contributions from Fe \textit{3d}-states (region \textbf{a}) and As \textit{4p}-states mixed with Fe \textit{3d,4s}-states (region \textbf{b}) are located at the bottom of the valence band (-2 to -5.5 eV). The alkali earth \textit{2s, 3s}-states (for Li and Na, respectively) provide a minimal but consistent contribution from 0 to -5.5 eV, these states hybridize with the Fe \textit{3d}-states.

To compare the XES spectra with the calculated DOS in a meaningful way, we separated the \textit{L}$_\mathit{3}$ band in the XES spectra into pseudo-Voigt components (of the form given in \ref{eqn:voigt}). As discussed, XES probes the partial occupied density of states; however the DOS features are broadened by both the instrumental resolution (Gaussian in nature) and the core-hole lifetime \cite{goodings69} (Lorentzian in nature). We only fit the \textit{L}$_\mathit{3}$ band since the \textit{L}$_\mathit{2}$ band is basically the same partial occupied density of states with poorer statistics. The position $\mu_i$, amplitude $A_i$, and Lorentz broadening $\Gamma_i$ for each component are determined by least-squares fitting. Here we set the Gaussian $\sigma$ parameter to the position $\mu_i$ divided by the instrumental resolving power (E/$\Delta$E), and kept a consistent mixing factor $\eta$ for all component peaks. We calculated best-fit curves with one to six pseudo-Voigt components and the quality of fit parameter $F^{\prime\prime} = \sqrt{\sum_x \left(f_{data}(x) - f_{fit}(x)\right)^2}$ was examined to determine the ``simplest best fit''. Since least-squares fitting requires initial estimates for the fitted parameters, fitting was conducted several times for a given number of pseudo-Voigt components with range of different initial positions $\mu_i$; this quantity is the hardest to fit accurately and is also the one we are most interested in. The initial estimates for $A_i$, $\Gamma_i$, and $\eta$ were set to 1.0, 1.0, and 0.5 respectively in all cases, these quantities are fitted quite well regardless of the initial estimate. We obtained consistent fits for several different initial estimates for $\mu_i$, indicating that the fit is unbiased by our choice for initial conditions. 
\begin{eqnarray}
\nonumber
f_V &=& A \left( \eta f_G + \left( 1 - \eta \right) f_L\right)\\
\nonumber
f_G &=& \frac{1}{\sqrt{2\pi}\sigma} \exp\left(-\frac{\left(x - \mu\right)^2}{2\sigma^2}\right)\\
f_L &=& \frac{1}{\pi} \left( \frac{\frac{\Gamma}{2}}{\left(x - \mu\right)^2 + \left(\frac{\Gamma}{2}\right)^2}\right)
\label{eqn:voigt}
\end{eqnarray}

\begin{figure}
\includegraphics[width=3in]{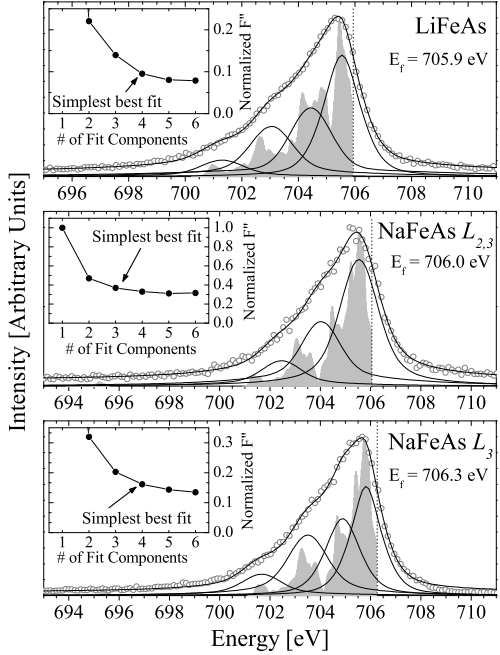}
\caption{\label{fig:curves}Comparison of calculated DOS and measured spectra. The top panel shows the \textit{L}$_{\mathit{3}}$ peak of the Fe \textit{L}$_{\mathit{2,3}}$ spectrum, the Fe \textit{3d} DOS, best-fit curve, and the 4 pseudo-Voigt components contributing to the best-fit curve for LiFeAs. The centre panel shows the \textit{L}$_{\mathit{3}}$ peak of the Fe \textit{L}$_{\mathit{2,3}}$ spectrum, the Fe \textit{3d} DOS, best-fit curve, and the 3 pseudo-Voigt components contributing to the best-fit curve for NaFeAs. The bottom panel shows the resonantly excited Fe \textit{L}$_{\mathit{3}}$ spectrum, the Fe \textit{3d} DOS, best-fit curve, and the 3 pseudo-Voigt components contributing to the best-fit curve for NaFeAs. The insets show the normalized fit parameter $F^{\prime\prime}$ for fits with different numbers of pseudo-Voigt components for each material. Note that the maximum $F^{\prime\prime}$ has been scaled to 1.0 in each case -- for example in LiFeAs the worst fit (with 1 pseudo-Voigt function) falls far outside the range of the plot. The estimated Fermi level is indicated in each plot.}
\end{figure}

For LiFeAs four or more pseudo-Voigt components produced fits with $F^{\prime\prime}$ values within 10\% of each other while the $F^{\prime\prime}$ for the best fit with 3 components was $\sim$25\% greater than the 4 component fit (see Figure \ref{fig:curves}, top panel, inset). Likewise, 3 or more pseudo-Voigt components produced a consistently good fit for NaFeAs, whereas the $F^{\prime\prime}$ for the best 2 component fit was $\sim$50\% greater than that of the 3 component fit. The fit for NaFeAs was not as good as for LiFeAs because the NaFeAs data had more noise. The fitted curve for LiFeAs matches the measured Fe \textit{L}$_{\mathit{3}}$ spectrum, and the components are located around the main features in the Fe \textit{3d} DOS calculation (Figure \ref{fig:curves}, top panel). Only the main Fe \textit{3d} DOS feature at 705.5 eV is sharp and isolated enough to potentially provide a pure pseudo-Voigt spectrum, so we expect the fitted components at lower energies to have higher amplitudes relative to the pseudo-Voigt at 705.5 eV than a comparison of the representative heights of the Fe \textit{3d} DOS features at the Fermi energy and elsewhere in the valence band would indicate. For example, the ratio between the heights of the Fe \textit{3d} DOS feature plotted at 705.5 eV and the feature plotted at 702.8 eV in the top panel of figure \ref{fig:curves} is much greater then the ratio between heights of the pseudo-Voigt component at 705.5 eV and the pseudo-Voigt component at 703.1 eV (see Table \ref{tbl:fit}), which is not unexpected since there are many other Fe \textit{3d} DOS features at $\sim$703 eV within the instrumental resolution of each other, and all of them would contribute to the spectrum, but would not be independently resolvable.

The fitted curve for NaFeAs matches the measured Fe \textit{L}$_{\mathit{3}}$ spectrum, but only one of the pseudo-Voigt components is in good agreement with a Fe \textit{3d} DOS feature (see Figure \ref{fig:curves}, centre panel). It is likely that the poorer quality of the NaFeAs XES spectrum compared to the LiFeAs XES spectrum is to blame here; it is possible to obtain a good fit using 4 pseudo-Voigt components fixed at 705.5 eV, 704.8 eV, 703.0 eV and 701.2 eV as the features in the Fe \textit{3d} DOS suggest, however our aim was to identify features in the DOS from the XES spectra without requiring input from calculations.

Resonant XES spectra usually have better signal-to-noise ratio, and since these materials do not show significant energy-loss features in the resonantly excited spectra, we can attempt to fit the resonantly excited Fe \textit{L}$_{\mathit{3}}$ spectrum by the method outlined above. In this case, 4 pseudo-Voigt components provide the optimal fit (Figure \ref{fig:curves}, inset in bottom panel) and these components give a much better agreement to the Fe \textit{3d} DOS than the 3-component non-resonant XES fitting did (see Figure \ref{fig:curves}, bottom panel). The fitted curve matches the measured spectrum except at the edge above 708 eV. This is not unexpected, since 708 eV is the excitation energy for this spectrum (see Figure \ref{fig:xes_summary}, top right panel) and the elastically scattered X-ray peak will distort the spectrum near 708 eV -- in particular by increasing the amplitude of the feature with less broadening (the elastic scatter has a purely Gaussian profile) then a pseudo-Voigt with parameters consistent with the remainder of the spectrum will be able to satisfy. The Fe \textit{L}$_{\mathit{3}}$ XES band from the other measured spectra (curves \textbf{b} and \textbf{c} in Figure \ref{fig:xes_summary} for both LiFeAs and NaFeAs) produced results consistent with those discussed above. In general, non-resonant XES should provide the best results if the signal-to-noise ratio is high enough, XES resonant with the \textit{L}$_\mathit{3}$ feature may be used if there are no energy-loss features or prominent scattering features.
\begin{table}
\begin{tabular}{ccccc}
&$\eta_i$ & $A_i$ & $\mu_i$ [eV] & $\Gamma_i$ [eV]\\
\hline
LiFeAs$_1$ & 0.47 & 1.48 & 705.5 & 1.34\\
LiFeAs$_2$ & & 1.44 & 704.4 & 9.12\\
LiFeAs$_3$ & & 0.96 & 703.1 & 5.24\\
LiFeAs$_4$ & & 0.29 & 701.3 & 4.15\\
NaFeAs$_1^N$ & 0.37 & 2.07 & 705.5  & 2.27\\
NaFeAs$_2^N$ & & 1.46 & 704.0 & 5.77\\
NaFeAs$_3^N$ & & 0.53 & 702.5 & 4.19\\
NaFeAs$_1^R$ & 0.39 & 1.28 & 705.8  & 1.25\\
NaFeAs$_2^R$ & & 1.05 & 704.9 & 1.61 \\
NaFeAs$_3^R$ & & 1.12 & 703.5 & 3.31\\
NaFeAs$_4^R$ & & 0.43 & 701.7 & 4.62\\
\hline
\end{tabular}
\caption{\label{tbl:fit}Fit results for the 4 pseudo-Voigt peaks for fitting the Fe \textit{L}$_{\mathit{3}}$ XES of LiFeAs and the 3 and 4 pseudo-Voigt peaks for fitting the nonresonant and resonant Fe \textit{L}$_{\mathit{3}}$ XES of NaFeAs (NaFeAs$_i^N$ and NaFeAs$_i^R$ respectively). Note there is only one mixing-factor $\eta$ for all components in the same spectrum.}
\end{table}

The Fermi levels in Figure \ref{fig:curves} were estimated by aligning the calculated DOS with the fitted pseudo-Voigt components, so the agreement between DOS and the first pseudo-Voigt component is manufactured. One method of estimating the Fermi level from emission spectra is to use the peak of the second derivative \cite{kurmaev08_2}. We have estimated Fermi levels of 705.9 eV, 706.0 eV, and 706.3 eV from aligning DOS and fit components and 706.5 eV, 706.8 eV, and 706.7 eV from the peak in the second derivative of the appropriate Fe \textit{L}$_{\mathit{3}}$ XES for for LiFeNa, non-resonant NaFeAs, and resonant NaFeAs, respectively. Since the main peak at 705.5 eV is so close to the Fermi level, the Fe \textit{3d} occupied states suffer an abrupt cut-off rather than a gradual decline at the Fermi level, and the \textit{L}$_{\mathit{3}}$ XES portions at higher energies than that are due only to spectral broadening. It is therefore not unexpected that the XES second-derivative estimates of the Fermi level are greater than the curve-fit alignment Fermi level estimate by a shift of roughly the instrumental energy resolution.

To summarize, we have studied the electronic structure of LiFeAs and NaFeAs via resonant and non-resonant XES spectra and DFT calculations using the generalized gradient approximation. We demonstrate that pseudo-Voigt curve fitting of XES measurements without knowledge of the electronic structure can give good agreement with prominent features in the calculated valence band. The results from DFT calculations and the comparison of Fe \textit{L}$_{\mathit{3}}$ FWHM and I(\textit{L}$_{\mathit{2}}$)/I(\textit{L}$_{\mathit{3}}$) ratio with standard reference compounds suggests the Fe \textit{3d} states are mostly itinerant in nature. The comparison of our results with previous studies on LaOFeAs and CaFe$_2$As$_2$ show that the main features of electronic structure found here are general for all studied FeAs-systems: Fe 3d-states dominate near the Fermi level and As 4p-states are concentrated at the bottom of the valence band. We conclude that all FeAs-systems studied herein are weakly or moderately correlated systems.

\section*{Acknowledgments}
We acknowledge support of the Research Council of the President of the Russian Federation (Grants NSH-1929.2008.2 and NSH-1941.2008.2), the Russian Science Foundation for Basic Research (Project 08-02-00148), the Natural Sciences and Engineering Research Council of Canada (NSERC), the Canada Research Chair program and the United Kingdom Engineering and Physical Sciences Research Council (EPSRC) for funding under grant EP/E025447.

\section*{References}
\bibliographystyle{unsrt}
\bibliography{lifeas}

\begin{thebibliography}{10}

\bibitem{pitcher08}
M.~J. Pitcher, D.~R. Parker, P.~Adamson, S.~J.~C. Herkelrath, A.~T. Boothroyd,
  and S.~J. Clarke.
\newblock {\em Chem.\ Commun.}, 45:5918, 2008.

\bibitem{tapp08}
J.~H. Tapp, Z.~Tang, B.~Lv, K.~Sasmal, B.~Lorenz, P.~C.~W. Chu, and A.M. Guloy.
\newblock {\em Phys.\ Rev.\ B}, 78:060505(R), 2008.

\bibitem{wang08}
X.~C. Wang, Q.~Q. Liu, Y.~X. Lv, W.~B. Gao, L.~X. Yang, R.~C. Yu, F.~Y. Li, and
  C.Q. Jin, 2008.
\newblock arXiv:cond-mat/0806.4688.

\bibitem{parker08}
D.~R. Parker, M.~J. Pitcher, and S.~J. Clarke, 2008.
\newblock arXiv:cont-mat/0810.3214.

\bibitem{chu09}
C.~W. Chu, F.~Chen, M.~Gooch, A.~M. Guloy, B.~Lorenz, B.~Lv, K.~Sasmal, Z.~J.
  Tang, J.~H. Tapp, and Y.~Y. Xue.
\newblock arXiv:cond-mat/0902.0806 (Physica C, in press).

\bibitem{jishi08}
R.~A. Jishi and H.~M. Alyahyaei, 2008.
\newblock arXiv:cond-mat/0812.1215v1.

\bibitem{kurmaev08}
E.~Z. Kurmaev, R.~G. Wilks, A.~Moewes, N.~A. Skorikov, Yu.~A. Izyumov, L.~D.
  Finkelstein, R.~H. Li, and X.~H. Chen.
\newblock {\em Phys.\ Rev.\ B}, 78:220503(R), 2008.

\bibitem{kurmaev09}
E.~Z. Kurmaev, J.~A. McLeod, A.~Buling, N.~A. Skorikov, A.~Moewes, M.~Neumann,
  M.~A. Korotin, Yu.~A. Izyumov, N.~Ni, and P.~C. Canfield, 2009.
\newblock arXiv:cond-mat/0902.1141.

\bibitem{jia95}
J.~J. Jia, T.~A. Callcott, J.~Yurkas, A.~W. Ellis, F.~J. Himpsel, M.~G. Samant,
  J.~St{\"{o}}hr, D.~L. Ederer, J.~A. Carlisle, E.~A. Hudson, L.~J. Terminello,
  D.~K. Shuh, and R.~C.~C. Perera.
\newblock {\em Rev.\ Sci.\ Instrum.}, 66:1394, 1995.

\bibitem{blaha01}
P.~Blaha, K.~Schwarz, G.~K.~H. Madsen, D.~Kvasnicka, and J.~Luitz.
\newblock {\em {\textbf{WIEN2k}},\ An\ Augmented\ Plane\ Wave\ +\ Local\
  Orbitals\ Program\ for\ Calculating\ Crystal\ Properties}.
\newblock Karlheinz Schwarz, Techn. Universit{\"{a}}t Wien, Austria, 2001.
\newblock {\textsc{ISBN}} 3-9501031-1-2.

\bibitem{perdew96}
J.~P Perdew, K.~Burke, and M.~Ernzerhof.
\newblock {\em Phys.\ Rev.\ Lett.}, 77:3865, 1996.

\bibitem{chen08}
G.~F. Chen, Z.~Li, D.~Wu, G.~Li, W.~Z. Hu, J.~Dong, P.~Zheng, J.~L. Luo, and
  N.~L. Wang.
\newblock {\em Phys. Rev. Lett.}, 100:247002, 2008.

\bibitem{ronning08}
F.~Ronning, T.~Klimczuk, E.~D. Bauer, H.~Volz, and J.~D. Thompson.
\newblock {\em J. Phys.: Condens. Matter}, 20:322201, 2008.

\bibitem{basinski55}
Z.~S. Basinski, W.~Hume-Rothery, and A.~L. Sutton.
\newblock {\em Proc. Royal Soc. London}, 229:459, 1955.

\bibitem{fjellvag96}
H.~Fjellvag, F.~Gronvold, S.~Stolen, and B.~C. Hauback.
\newblock {\em J. Solid State Chem.}, 124:52, 1996.

\bibitem{galakhov97}
V.~R. Galakhov, A.~I. Poteryaev, E.~Z. Kurmaev, V.~I. Anisimov, St. Bartkowski,
  M.~Neumann, Z.~W. Lu, B.~M. Klein, and T.-R. Zhao.
\newblock {\em Phys.\ Rev.\ B}, 56:4584, 1997.

\bibitem{gao06}
X.~Gao, D.~Qi, S.~C. Tan, A.~T.~S. Wee, X.~Yu, and H.~O. Moser.
\newblock {\em J.\ Electr.\ Spectr.\ Relat.\ Phenom.}, 151:199, 2006.

\bibitem{kurmaev05}
E.~Z. Kurmaev, A.~L. Ankudinov, J.~J. Rehr, L.~D. Finkelstein, P.~F. Karimov,
  and A.~Moewes.
\newblock {\em J.\ Electr.\ Spectr.\ Relat.\ Phenom.}, 148:1, 2005.

\bibitem{goodings69}
D.~A. Goodings and R.~Harris.
\newblock {\em J.\ Phys.\ C}, 2(2):1808, 1969.

\bibitem{kurmaev08_2}
E.~Z. Kurmaev, R.~G. Wilks, A.~Moewes, L.~D. Finkelstein, S.~N. Shamin, and
  J.~Kun{\u{e}s}.
\newblock {\em Phys.\ Rev.\ B}, 77:165127, 2008.

\end{thebibliography}

\end{document}